\documentclass[conference]{IEEEtran}
\IEEEoverridecommandlockouts

\usepackage{cite}
\usepackage{amsmath,amssymb,amsfonts}
\usepackage{algorithmic}
\usepackage{graphicx}
\usepackage{textcomp}
\usepackage{xcolor}
\usepackage{graphicx}
\usepackage{url}
\usepackage[caption=false,font=footnotesize]{subfig}
\def\BibTeX{{\rm B\kern-.05em{\sc i\kern-.025em b}\kern-.08em
    T\kern-.1667em\lower.7ex\hbox{E}\kern-.125emX}}
\begin{document}

\title{Large Data Acquisition and Analytics at Synchrotron Radiation Facilities}
\newcommand{\aashish}[1]{\textcolor{blue}{[Aashish: #1]}}
\newcommand{\amy}[1]{\textcolor{red}{[Amy: #1]}}
\author{
\IEEEauthorblockN{Aashish Panta}
\IEEEauthorblockA{\textit{University of Utah}\\
Salt Lake City, USA}
\and
\IEEEauthorblockN{Giorgio Scorzelli}
\IEEEauthorblockA{\textit{University of Utah}\\
Salt Lake City, USA}
\and
\IEEEauthorblockN{Amy A. Gooch}
\IEEEauthorblockA{\textit{ViSOAR}\\
Salt Lake City, USA}
\and
\IEEEauthorblockN{Werner Sun}
\IEEEauthorblockA{\textit{CHESS}\\
Ithaca, USA}
\and
\IEEEauthorblockN{Katherine S. Shanks}
\IEEEauthorblockA{\textit{CHESS}\\
Ithaca, USA}
\and 
\IEEEauthorblockN{Suchismita Sarker}
\IEEEauthorblockA{\textit{CHESS}\\
Ithaca, USA}
\and
\IEEEauthorblockN{Devin Bougie}
\IEEEauthorblockA{\textit{CHESS}\\
Ithaca, USA}
\and 
\IEEEauthorblockN{Keara Soloway}
\IEEEauthorblockA{\textit{CHESS}\\
Ithaca, USA}
\and  
\IEEEauthorblockN{Rolf Verberg}
\IEEEauthorblockA{\textit{CHESS}\\
Ithaca, USA}
\and
\IEEEauthorblockN{Tracy Berman}
\IEEEauthorblockA{\textit{University of Michigan}\\
Ann Arbor, USA}
\and
\IEEEauthorblockN{Glenn Tarcea}
\IEEEauthorblockA{\textit{University of Michigan}\\
Ann Arbor, USA}
\and
\IEEEauthorblockN{John Allison}
\IEEEauthorblockA{\textit{University of Michigan}\\
Ann Arbor, USA}
\and
\IEEEauthorblockN{Michela Taufer}
\IEEEauthorblockA{\textit{University of Tennessee, Knoxville}\\
Knoxville, USA}
\and
\IEEEauthorblockN{Valerio Pascucci}
\IEEEauthorblockA{\textit{University of Utah}\\
Salt Lake City, USA}
}

\maketitle

\begin{abstract} 
 Synchrotron facilities like the Cornell High Energy Synchrotron Source (CHESS) generate massive data volumes from complex beamline experiments, but face challenges such as limited access time, the need for on-site experiment monitoring, and managing terabytes of data per user group. We present the design, deployment, and evaluation of a framework that addresses CHESS’s data acquisition and management issues. Deployed on a secure CHESS server, our system provides real-time, web-based tools for remote experiment monitoring and data quality assessment, improving operational efficiency. Implemented across three beamlines (ID3A, ID3B, ID4B), the framework managed 50–100 TB of data and over 10 million files in late 2024. Testing with 43 research groups and 86 dashboards showed reduced overhead, improved accessibility, and streamlined data workflows. Our paper highlights the development, deployment, and evaluation of our framework and its transformative impact on synchrotron data acquisition.

\end{abstract}

\begin{IEEEkeywords}
 Large Data Acquisition, Remote Monitoring, Synchrotron Radiation Facilities, Scientific Workflows, Beamline, Dashboard, Web-based analytics,  Decision Making\end{IEEEkeywords}

\section{Introduction and Motivation}
Large synchrotron radiation facilities like the Cornell High Energy Synchrotron Source (CHESS) are vital hubs for advancing research across domains including quantum physics, structural biology, and materials science. Using powerful X-ray beamlines, synchrotron facilities enable cutting-edge experiments that generate vast data but also face logistical and technical challenges during acquisition. CHESS, supported by the National Science Foundation, National Institutes of Health, and Air Force Research Laboratory, features seven state-of-the-art beamlines, each hosting 1–10 user groups weekly. Serving ~1,300 users annually, CHESS generates about 400 TB of raw data per year, stored across ~2 PB of disk capacity and a magnetic tape library. Each beamline is a complex system of specialized devices with tightly scheduled data acquisition sessions (``beamtimes"), requiring researchers to reserve time months in advance. With limited access periods of 1–6 days per group, efficient use of beamtime is critical to achieving experimental objectives. Similar challenges exists at nearly all large-scale X-ray and neutron facilities, making scalable data management solutions broadly relevant.

Before 2020, most CHESS users traveled to Ithaca, New York, for their beamtimes. During the COVID-19 pandemic, CHESS deployed a suite of remote access tools that remain in use today, centered on SSH and the NoMachine Enterprise product family~\cite{NoMachineEnterprise2025} for remote access to physical displays and virtual desktops. 

\begin{figure}[t!]
    \centering
    \includegraphics[width=0.99\linewidth]{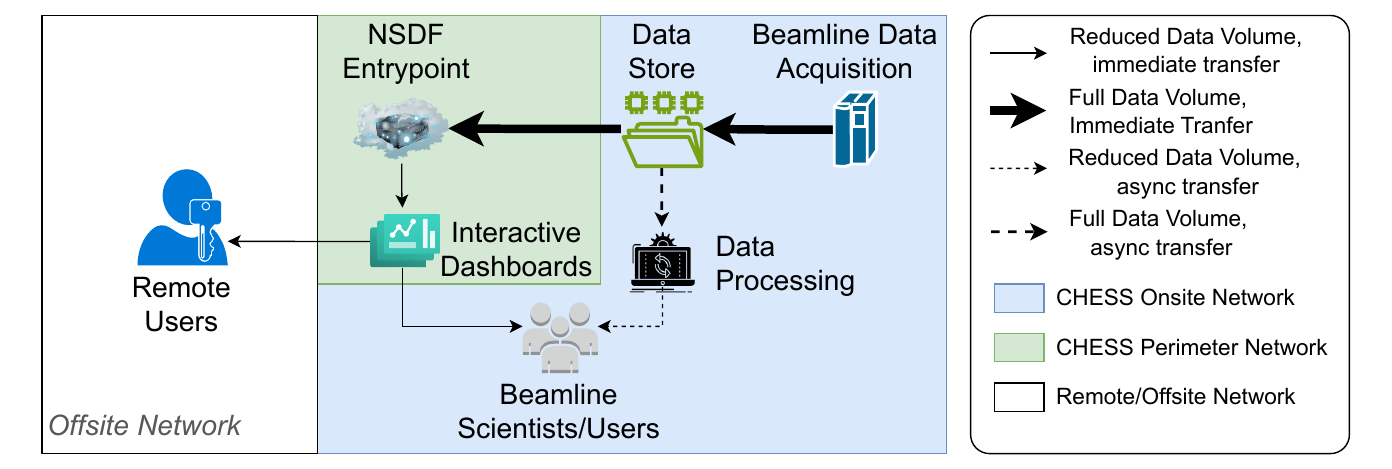}
    \caption{ Our remote data acquisition and evaluation framework. Traditionally, scientists interact with the physical display of the beamline station computer to collect and monitor data. With our framework’s EntryPoints deployment, data is now acquired, visualized, accessed, and able to be analyzed remotely, significantly enhancing experimental flexibility and efficiency.}
    \label{fig:overview}
     \vspace{-4mm}
\end{figure}

Historically, when not on-site, scientists logged in remotely to CHESS beamline control and analysis computers to monitor experiments and assess data quality. While functional, this setup was inconvenient, prone to inefficiencies, and limited in interpretation capabilities. Researchers lacked tools to monitor data generation rates, detect corrupted files, or evaluate dataset integrity in real time, making it difficult to manage limited beam time. It also required considerable familiarity with CHESS infrastructure, remote access tools, and analysis frameworks, often delaying troubleshooting during critical experiments. The sheer scale of data produced—several terabytes daily for a single user group—further compounded challenges in management, validation, and quality assurance.

To address these challenges, we developed and deployed an innovative web-based framework to improve efficiency, accessibility, and transparency of data acquisition workflows at CHESS. Currently tailored to three of seven beamlines—ID3A, ID3B, and ID4B (Section~\ref{sec:deploy_beamline})—the framework integrates seamlessly into existing pipelines, enabling scientists to monitor experiments remotely and in real time.

The contributions of this paper are as follows.

\begin{itemize}
\setlength\itemsep{1mm}
    \item \textbf{High-Throughput Data Acquisition and Access Architecture:} We introduce a modular framework that supports secure scalable access to high-rate CHESS beamline acquisition data. The system enables remote monitoring and analysis of multi-terabyte experiments by connecting on-site acquisition systems with external collaborators through dedicated EntryPoints, overcoming long-standing limitations imposed by secure facility networks.

    \item \textbf{Real-Time Data Validation and Streaming Analytics:} We develop automated pipelines for near real-time file integrity checks, metadata extraction, error detection, throughput tracking, on-the-fly visualization and dataset restructuring. These capabilities surface anomalies in acquisition patterns and data quality early, enabling users to intervene before significant beamtime is lost.  

    \item \textbf{Scalable Workflow Optimization for Data-Intensive Experiments:} By automating the ingestion and transformation of heterogeneous data formats and consolidating metadata across hundreds of thousands of files, the framework reduces operational overhead and supports adaptive, data-driven decision making during complex in situ experiments. Our framework allows multiple research groups to simultaneously evaluate their progress on the fly, enabling critical real-time tuning of experimental parameters and data-driven decision making for complex in situ experiments. 

     \item \textbf{Domain Expert Assessment and Validation:} Our framework, consisting of an acquisition dashboard and a data evaluation dashboard for each of the 43 research groups over two years, continues to be iteratively developed and assessed by the CHESS staff scientists and user groups. Our paper presents informal validation from CHESS staff scientists and users, highlighting the utility of existing features and directions for ongoing development.  

\end{itemize}

CHESS has a long and fruitful history of pursuing exploratory research and development projects with domain science users. CHESS personnel routinely collaborate with users to develop new X-ray science techniques that can be replicated at other facilities. Often, this process requires CHESS to accommodate the bespoke equipment, computers, or software that users wish to test during an experiment.
The interdisciplinary project described here follows this model of close collaboration. In the discovery phase, we established a shared vocabulary between X-ray and computer science concepts and identified urgent needs addressable by real-time visualization. CHESS scientists provide feedback at each stage, especially during dashboard development for data acquisition and analytics, ensuring that their design was use-driven and readily adopted.

Our collaboration continues to refine the dashboards based on real-world experience. The framework has transformed workflows at CHESS, addressing long-standing data acquisition challenges and proving to be a game changer in optimizing beamline operations and experimental outcomes.

\begin{figure*}[!t]
  \centering
  \subfloat[One of our widgets for the statistics dashboard.\label{fig:timing_down}]{
      \includegraphics[height=.28\linewidth]{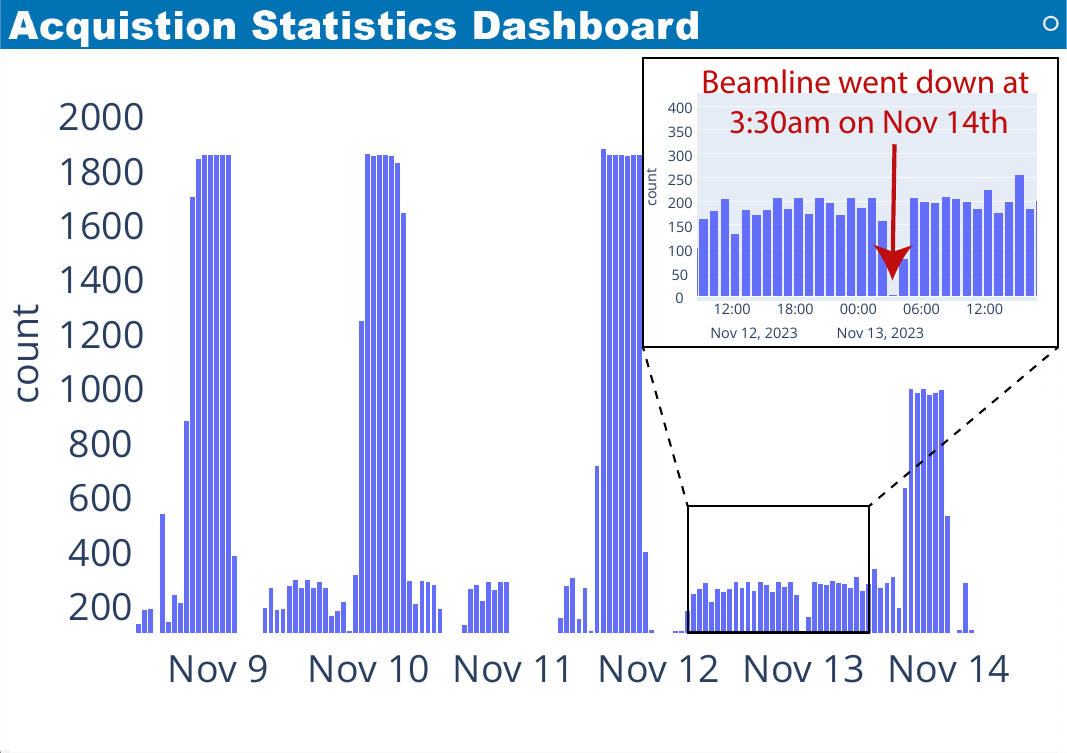}
  }\hfill
  \subfloat[Probe dashboard.\label{fig:probe_with_missing_data}]{
      \includegraphics[height=.28\linewidth]{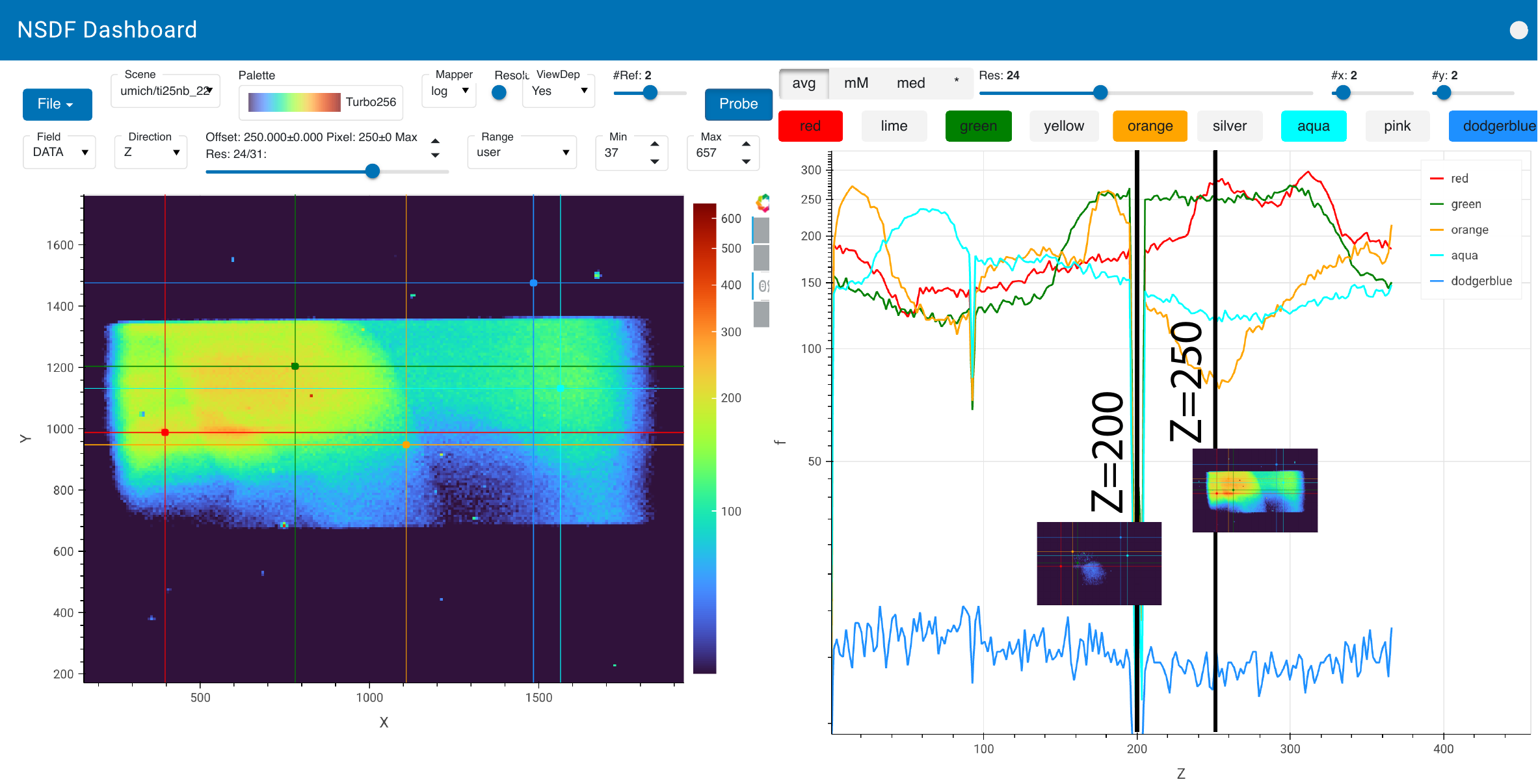}
  }
  \caption{Our web-based  framework revolutionizes synchrotron data acquisition and analysis by enabling remote monitoring, real-time data quality assessment, and improved operational efficiency, significantly enhancing scientific productivity and on-site or remote collaboration at large-scale facilities like CHESS. Our acquisition statistics dashboard features widgets that reveal previously hidden metadata to beamline user groups and enable real-time progress monitoring for on- and off-site user groups. (a) shows a timeline widget that enables zooming into specific event ranges to identify unexpected drops in file counts, highlighting interruptions in data collection, such as the beam outage at 3:30 AM on November 14, 2023. (b) shows our probe dashboard, comparing individual radiographs within a rotation series (left) and 5-probe view (right) for an arbitrary specimen. Each probe is configured to sample minimum and maximum in an 8x8 pixel sample. The main view shows the XY-plane at $Z=250$. The graph includes insets of the slices for illustrative purposes, including $Z=250$, where the green probe reaches a local minimum, and $Z=200$, where the image is lost due to intermittent intensity fluctuations.}
  \label{fig:teaser}
  \vspace{-4mm}
\end{figure*}
 
 \section{Background}
\subsection{X-Ray Specific Data Visualization Tools}
Modern synchrotron X-ray facilities such as CHESS require specialized visualization tools to deal with the unique challenges of beamline instrumentation and data analysis. Beamline scientists and users have historically relied on a combination of general-purpose and domain-specific tools, including Jupyter notebooks tailored to specific workflows and maintained by individual researchers, ImageJ for basic image processing~\cite{Schneider2012}, NeXpy for NeXus data file exploration~\cite{Klabunde2022}, HEXRD~\cite{HEXRD2021} for diffraction data evaluation and analysis, and HyperSpy for multidimensional data analysis~\cite{Hyperspy2017}, among others. These tools often require local installation and specialized expertise, leading to fragmented workflows across experimental stations.
Jerome et al. \cite{10386508} presented methods to reduce synchrotron micro-CT data from gigabytes to megabytes and provided visual previews using server-side processing and histogram filtering approaches but does not address issues with larger terascale synchrotron datasets. Chae et al.\cite{9006048} demonstrated how interactive visual analytics can make deep-learning embeddings of terabyte-scale molecular dynamics simulations interpretable, highlighting approaches relevant for handling and visualizing large, complex datasets such as those generated at synchrotron facilities. Aggour et al.\cite{9006495} designed a federated big data platform that integrates multimodal storage, analytics, and visualization—an approach relevant to large-scale experimental data workflows.

Recent advances in web-based visualization address some accessibility challenges. For example, PGMweb~\cite{Wang2025} enables browser-based simulation of X-ray beam paths through monochromators, eliminating installation requirements. The CHESS facility’s experience reflects these broader trends: in the absence of unified solutions, researchers still rely on a patchwork of bespoke scripts and tools, leading to maintenance challenges and skill silos, as techniques learned at one beamline often fail to transfer to others.

\subsection{Large-Scale Web-based Visualizations}
Web-based visualization libraries like D3~\cite{6064996} manipulate the \emph{document object model (DOM)}, while WebGL-powered tools such as Deck.gl~\cite{wang2019deck} and Luma.gl enable GPU-accelerated 3D rendering. Recent work has advanced browser-based visualization of large datasets: Usher et al.\cite{usher2020interactive} visualized terabytes of scientific data using block-compressed isosurfaces; Alder et al.\cite{alder2015web} built the USGS Climate Change Viewer; and Walker et al.\cite{w12102928} highlighted latency challenges in geospatial data. Server-side approaches like ParaViewWeb\cite{jourdain2011paraviewweb} and microservice frameworks~\cite{raji2018scientific} address network limitations.

Frameworks such as OpenVisus~\cite{pascucci2012visus} and OpenVisusPy restructure data via multiresolution space-filling curves~\cite{summa2011interactive}, enabling interactive exploration and post-hoc queries for petascale datasets. Its cache-oblivious methods~\cite{Venkat2016,Lindstrom2005} optimize storage for modern hardware, while progressive encoding of resolution and precision~\cite{Hoang21,11044452} reduces data movement. OpenVisus dashboards~\cite{Aashish_LDAV_2024} further allow progressive in-browser visualization without large downloads.

Browser-based visualization enhances accessibility and efficiency across domains. Large-scale initiatives support this trend:  Open Science Grid delivers distributed high-throughput computing (1.1M+ daily core hours)\cite{OSG2025};  Open Science Data Federation provides a global caching network for petascale workflows\cite{OSDFCache2024,ArgonneOSG2024};  National Data Platform integrates semantic tools with Jupyter for FAIR-compliant analysis~\cite{WiFireOntology2024,NDPNSF2023}; and NSDF offers layered services combining browser interfaces, a 1.5B+ record catalog, and OSG backbone~\cite{NSDF2023,SDSCNSDF2023}. We employ NSDF’s EntryPoint to integrate OpenVisus into our data management and visualization framework.

\section{Interdisciplinary Collaboration} \label{sec:collab}
The design and deployment of our framework required tightly coordinated expertise from beamline scientists, computer scientists, and IT engineers. This collaboration ensured that the system addressed the fundamental big data challenges present at CHESS, including terascale data streams, heterogeneous detector outputs, complex metadata relationships, and the need for automated, low-latency processing. The resulting framework reflects both scientific requirements and the operational constraints of a large-scale experimental facility.

From the early stages of design, beamline scientists helped define the data handling and analysis capabilities needed to support real experiments. These included reliable streaming of detector output, real-time verification of file integrity, consistent capture of experiment metadata, and rapid feedback that allows users to adjust scanning parameters during limited beamtime.  Computer scientists and system researchers translated these scientific needs into concrete  requirements. These included scalable data workflows capable of handling millions of pixels per frame, processing pipelines that could convert and restructure large datasets without interrupting live acquisition, and distributed storage layouts that could sustain both high write throughput and rapid downstream access. Supporting the heterogeneity of CHESS data was a central challenge during the development phase. Far-field diffraction detectors routinely produce more than 11 million pixels per frame in HDF5, while radiographs and near-field diffraction patterns arrive as multi-gigabyte TIFF stacks with around 4 million pixels per frame. Addressing these differences required the development of automated conversion pipelines, optimized chunking strategies, and multiresolution data representations suitable for interactive or batch analysis.

Early deployments on the ID3A beamline validated the system under live experimental loads and demonstrated its utility for detecting acquisition problems, monitoring detector behavior, and verifying scan completeness. These tests motivated the introduction of additional features such as real-time file-generation statistics and improved performance for multiresolution visualization. Subsequent deployments on ID3B and ID4B confirmed that the design generalizes across detectors, modalities, and data rates. The continuous feedback from beamline scientists ensured that the system remained aligned with scientific workflows while computer scientists iteratively refined algorithms, data flows, and interface components.

This collaboration produced a framework that is technically robust, scientifically grounded, and practical for use under time-critical conditions. The following sections describe the resulting components: the CHESS EntryPoint for secure high-throughput data access, the data acquisition dashboard for monitoring collection statistics, and the data evaluation dashboard for large 2D and 3D datasets.
 
\begin{figure*}[!t]
    \centering
    \subfloat[Percentage of files per extension.\label{fig:files_per_extension}]{
        \includegraphics[width=0.31\linewidth]{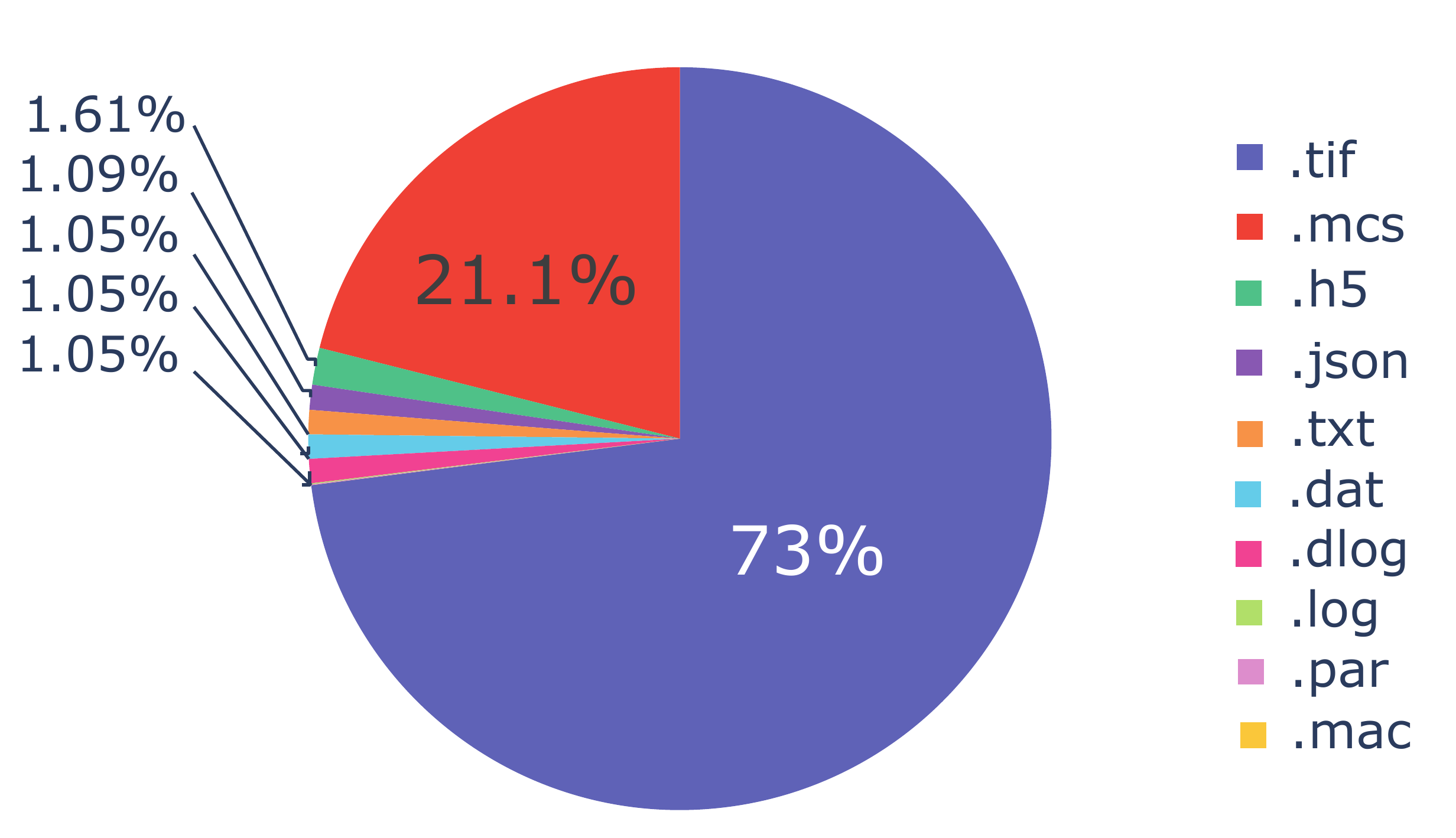}
    }\hfill
    \subfloat[Beamline experiment timeline.\label{fig:beamline_timeline}]{
        \includegraphics[width=0.31\linewidth]{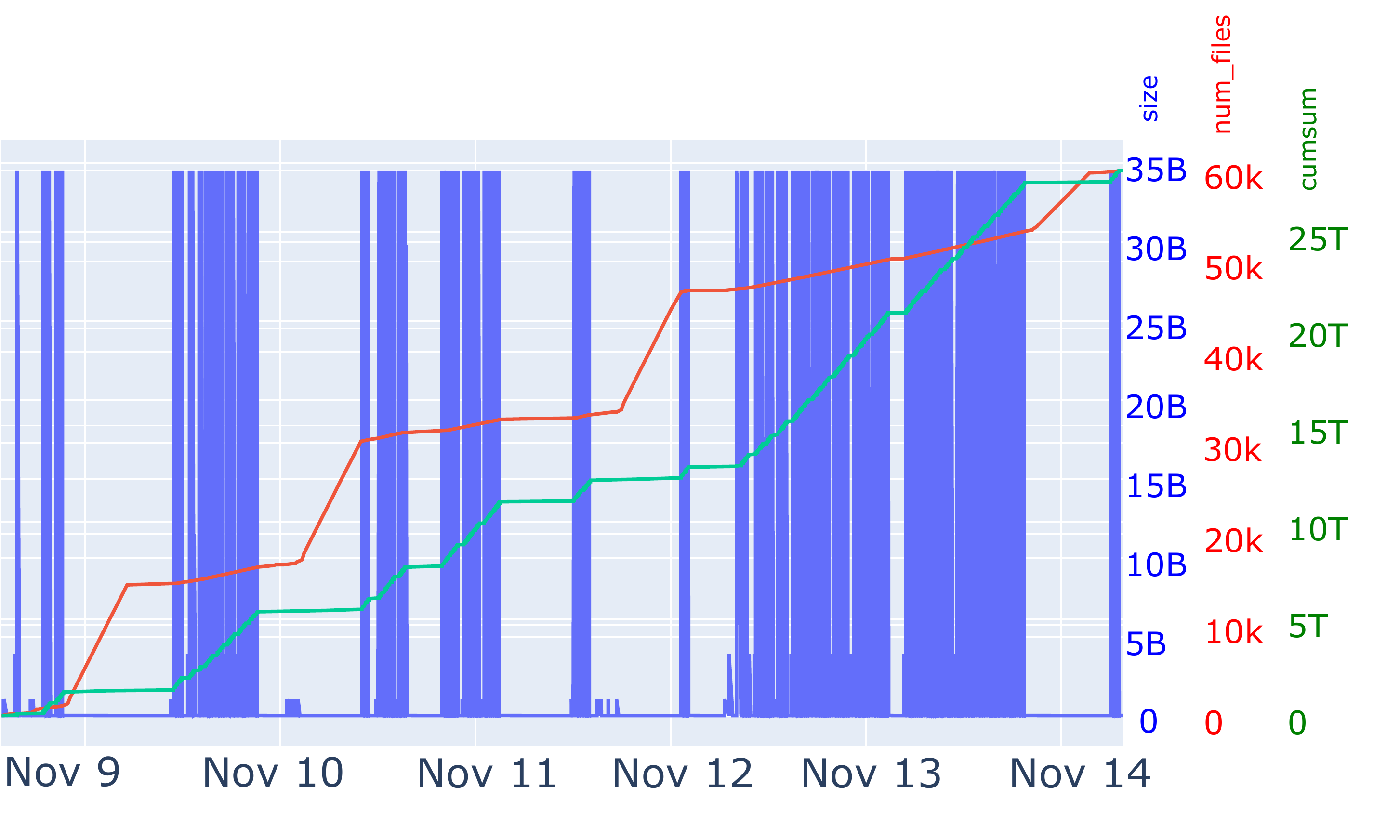}
    }\hfill
    \subfloat[Distribution of far-field (ff) and near-field (nf) HEDM scans.\label{fig:far_and_near_field}]{
        \includegraphics[width=0.31\linewidth]{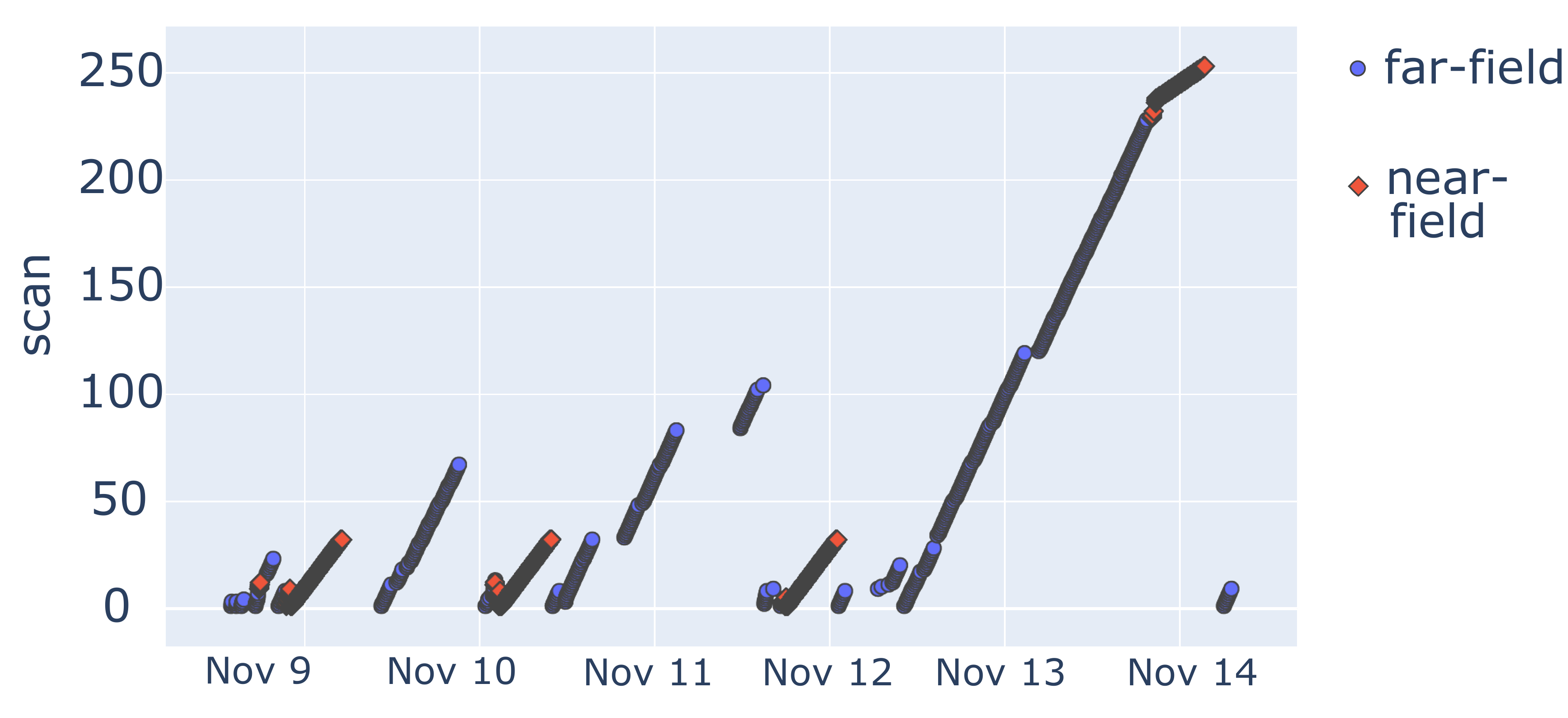}
    }
    \caption{Example statistics dashboard: (a) file extension distribution, (b) experiment timeline, and (c) timeline of far-field (ff) and near-field (nf) scans.}
    \label{fig:stats_dashboard}
    \vspace{-4mm}
\end{figure*}

In November 2023, members of the computer science team traveled to CHESS for the first major deployment during scheduled user experiments. The visit provided first-hand insight into beamline workflows and opportunities where real-time monitoring could enhance productivity, while testing the data pipeline, validating large-scale acquisition, and integrating with beamline infrastructure. This preliminary phase was critical to validate the framework’s ability to handle large-scale data acquisition, ensuring seamless integration with the beamline infrastructure before full-scale deployment.

The framework was first integrated into beamline ID3A, where real-time visualization exceeded expectations by allowing researchers to monitor progress remotely and reduce manual checks. We also recognized that file-generation statistics would further aid developers and users, leading to iterative widget improvements and reimplementation with newer libraries (e.g., replacing Bokeh with Panel).

Integration quickly showed workflow benefits, as scientists could identify and correct data quality issues to maximize limited beamtime. Feedback from users guided refinements before extending the deployment to ID3B and ID4B. Our collaboration demonstrated the immense value of combining expertise from different disciplines. The beamline scientists contributed their deep understanding of experimental setups, instrument behavior, and scientific data interpretation, while the computer scientists on the team brought expertise in data management, visualization, and computational workflows. Our collaboration ensured that the deployed framework was technically sophisticated and highly practical for end users. Our direct involvement with beamline scientists throughout the development and testing process created a framework that seamlessly fits into their workflow.

In the next sections, we describe the framework’s components: 1) CHESS EntryPoint for secure connections inside and outside CHESS, 2) the data acquisition dashboard for collection statistics, and 3) the data evaluation dashboard for large 2D/3D datasets.

\section{Our Framework}

Our data processing framework, depicted in Figure~\ref{fig:overview}, was designed to support the big data challenges encountered during high-throughput synchrotron experiments at facilities like CHESS. The system enables real-time monitoring, automated data restructuring, and remote access for distributed research teams, but its core contribution lies in the underlying dataflow and algorithms that make these capabilities possible under CHESS-scale workloads. Beamline experiments frequently generate tens of terabytes over a few days, comprising hundreds of thousands of files in formats such as NeXus, HDF5, CBF, TIFF, YAML, and JSON. Managing these heterogeneous and rapidly produced datasets in near real time requires a pipeline that can sustain high ingestion rates, extract and reconcile metadata across formats, detect corruption and missing frames, and reorganize raw detector data into multi-resolution layouts suitable for interactive exploration. These tasks must run continuously while acquisition is ongoing, which introduces practical difficulties such as partially written files, asynchronous metadata updates, inconsistent directory structures, and variable filesystem latencies. Our framework addresses these issues through a set of automated, fault-tolerant routines that operate incrementally, ensuring that live data can be inspected without interfering with the actual beamline operations. The dashboards presented to users are built on top of this infrastructure and serve primarily as interfaces to the underlying processing pipelines that enable real-time, distributed, and secure access to data acquisition.

\subsection{NSDF EntryPoint}\label{sec:entrypoint} 

One of the key challenges in integrating the dashboard framework with the CHESS beamlines was the facility’s secure internal network, which restricted external access to experimental data. While these security measures are essential for protecting sensitive research data and maintaining system integrity, they pose a significant limitation for enabling real-time remote monitoring and data evaluation. Researchers outside the facility, including those from collaborating institutions, needed remote access to CHESS computing resources to retrieve or analyze data during active acquisition sessions, increasing the complexity required and reducing the potential for distributed collaboration and timely decision-making. 

To address this challenge, we deployed an NSDF  EntryPoint within the CHESS perimeter network~\cite{NSDF_2023_Taufer}, illustrated in Figure~\ref{fig:overview}. The NSDF EntryPoint functioned as an intermediary data access gateway, enabling controlled and secure external access to the beamline data while adhering to facility security protocols. Due to the modular and well-engineered CHESS environment, we could slot the NSDF EntryPoint into the CHESS environment and access the data generated by the beamlines.
SPEC~\cite{spec}, a widely used instrument control and data acquisition software, controls the beamlines. The developers at CHESS were able to add extensions to SPEC that notify the NSDF software when the collection runs complete.  Once the collection runs are completed, it triggers a series of data processing scripts that include, but are not limited to,  extracting logs from metadata, checking for corrupted files, gathering statistics about the data itself, and making it ready for conversion to the data layout we can work with. 

As illustrated in Figure~\ref{fig:overview}, the CHESS EntryPoint is deployed inside the CHESS infrastructure, and acts as a bridge between the CHESS internal network and the NSDF network.  The CHESS EntryPoint consists of a 10Gb connection to the internal CHESS network, which it uses to stream data from the central CHESS Data Acquisition System (DAQ), a highly available and redundant enterprise storage cluster. The CHESS EntryPoint streams data to the local buffer on the NSDF EntryPoint, a 131 TB software RAID6 array composed of twelve 15 TB SATA drives.  Dashboards are connected to the data that is served to remote CHESS users.  The data is streamed to the NSDF infrastructure over a dedicated 25Gb perimeter network connection to Internet2.

The deployment process involved configuring a secure data pipeline that allowed experimental data to be streamed from the beamline instruments to the NSDF EntryPoint. From there, researchers with appropriate credentials could retrieve the data through authenticated channels. This setup ensured that security policies were maintained while simultaneously providing the necessary external access to data, enhancing the dashboard’s remote monitoring capabilities, facilitating real-time collaboration, and paving the way for future advancements in synchrotron data accessibility.

Remote access to CHESS dashboards, which run only on localhost, is provided through an NGINX forward proxy, ensuring secure, monitored, and logged connections. Static HTML files with beamtime statistics are served via an Apache HTTPD server, also proxied through NGINX for off-site access. Off-site connections are limited to ports 443 (HTTPS) and 22 (SSH) through the CHESS firewall. Access to NGINX, Apache, and NSDF dashboards is further restricted via CHESS Active Directory, granting availability only to scientists and users associated with specific beamtime experiments.

  \begin{figure*}[!phbt]
    \centering
    \includegraphics[width=0.45\linewidth]{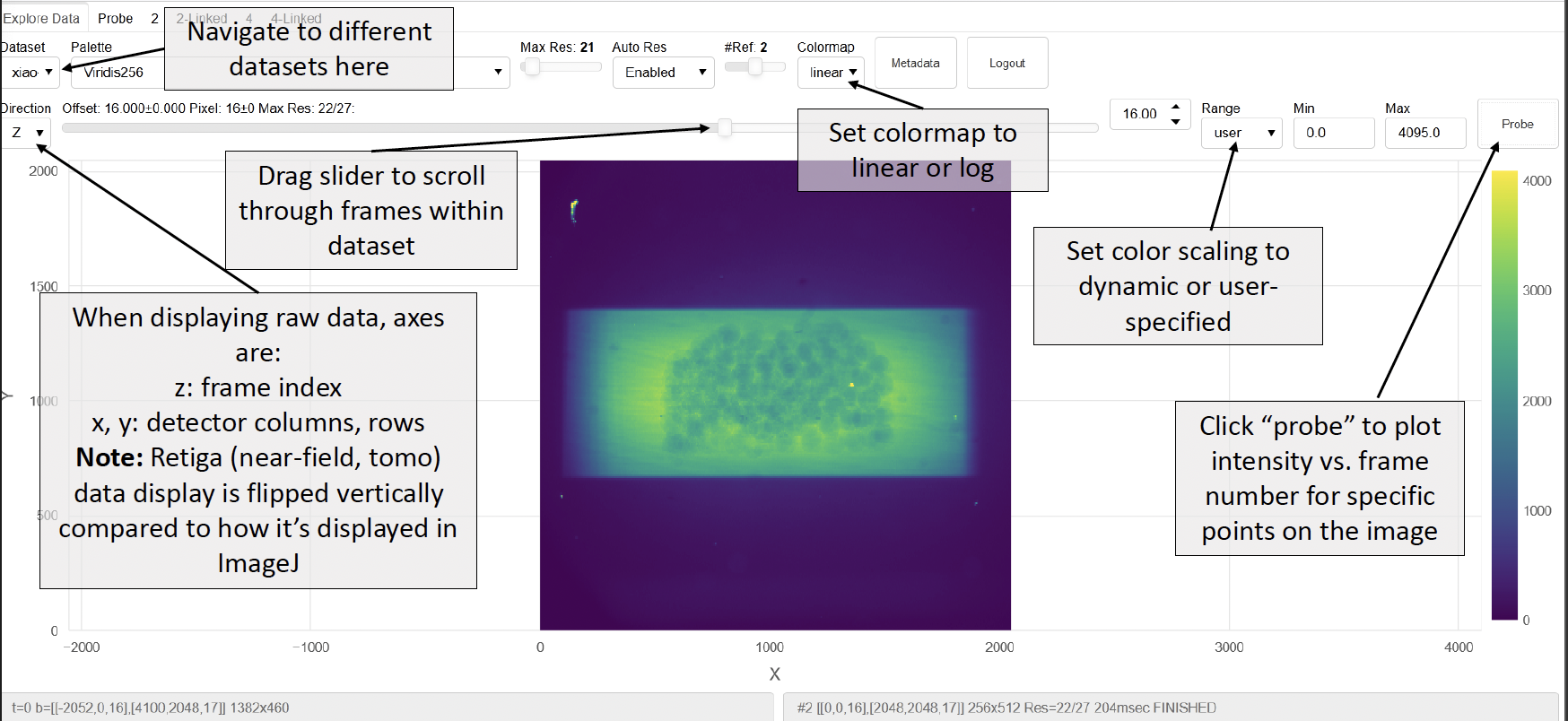}
    \hfill
    \includegraphics[width=0.45\linewidth]{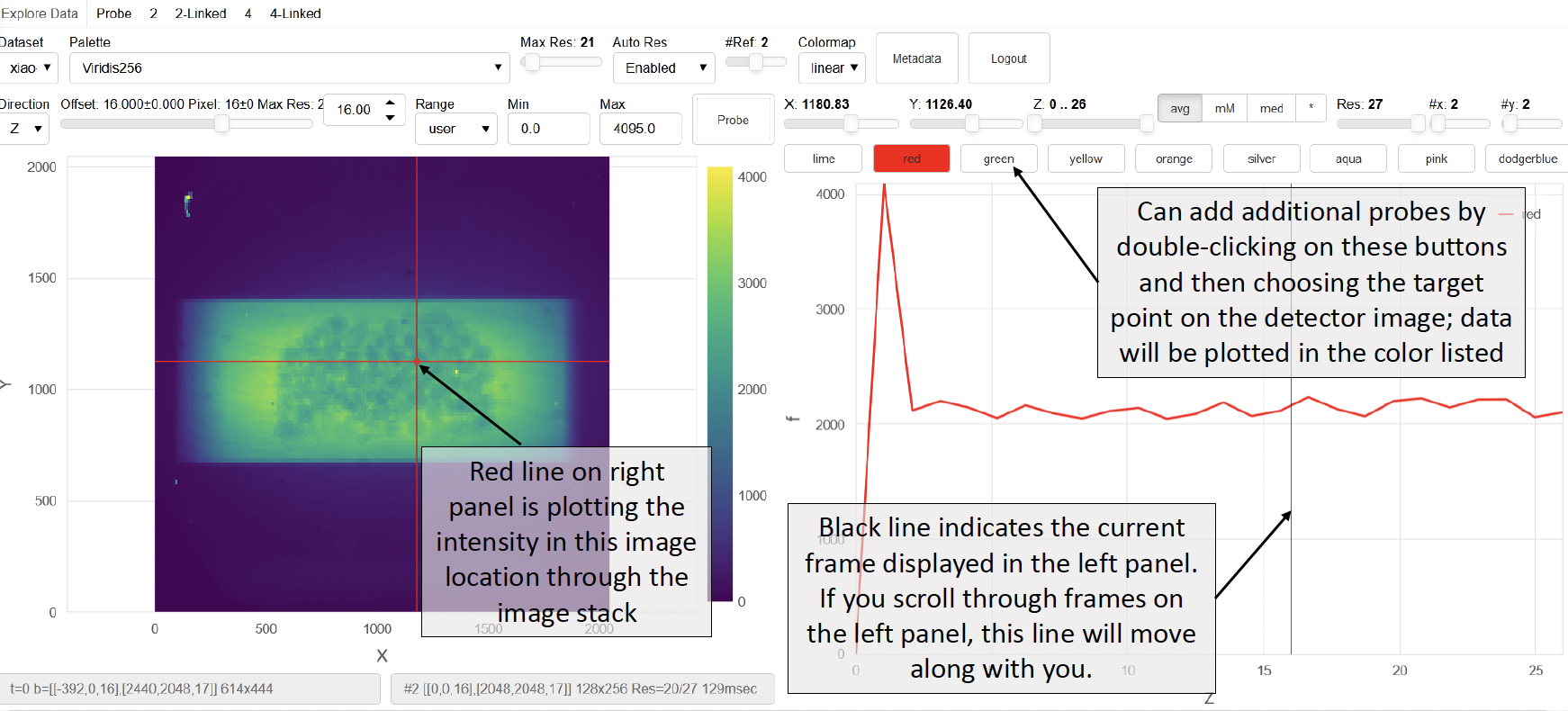}
    \caption{ The data evaluation and visualization dashboard, iteratively developed with feedback from CHESS staff scientists,  for evaluating data by slice of the volume (left). Advanced features in the probe panel of the dashboard (right) enable users to examine data through the volume at key points of interest.  One can dynamically change the view by moving the graph line or points in the slice view.  The probes can also be changed to provide average, maximum, or minimum within a window of [1,8] pixels in either direction.
 }
    \label{fig:dash_explained}
        \vspace{-4mm}
\end{figure*}

\subsection{Data Acquisition Dashboard for Beamline Statistics}\label{sec:dataAcq} 

Our framework includes data acquisition dashboards that visualize the progress of data collection in real-time and integrate seamlessly into the CHESS pipeline, supporting monitoring and quality control.

Custom Jupyter Notebooks and dashboards solve the need for users to monitor experiment progress constantly.  Our dashboards automatically populate for each acquisition run and facilitate a quick preview of the results without waiting for the entire experiment to complete.  Our approach extracts and integrates metadata embedded within experimental datasets previously inaccessible to end users. The metadata, derived from multiple file formats including Nexus, YAML, JSON, HDF5, and TIFF, as well as from system-level “storage” attributes (e.g., file creation and modification dates, file counts, and aggregate file sizes), was initially challenging to process due to its complexity and varied origins. By collaborating closely with beamline scientists, we contextualize these diverse data streams, ultimately leveraging NSDF to provide users with direct, frictionless access to these critical metadata from the interactive dashboards.
 
The acquisition statistics dashboard processes this metadata and allows interactive exploration. Because of the CHESS data management system's and our modular designs, we can exchange components, rewrite systems, and optimize transformations. During acquisition, statistics refresh continuously with a goal of refreshing every ten seconds, though in practice updates can take longer due to slow filesystem scraping. Once acquisition is complete, statistics remain static, except for one beamline where post-processing triggers updates every 30–60 minutes. The interface uses Plotly Express, selected over Bokeh and Panel for its performance on million-point 1D signals and features such as zooming and PNG export. Data processing relies on xarray, pandas, and h5py. Finalized workflows are converted from Jupyter notebooks to static HTML via jupyter nbconvert and served through Apache HTTPD, eliminating the need for Jupyter servers.

We configure each acquisition dashboard for the needs of the user group (Figure~\ref{fig:stats_dashboard}).  General widgets include a pie chart of the number of files per extension, a bar chart of the total size per extension, and data acquisition timeline. The beamlines for this deployment generate data in NeXus, HDF5, CBF, and TIFF file formats. Each beamtime produces up to 100 terabytes of raw data, with hundreds of thousands of files.  We also include timelines with the number of files uploaded and the accumulated sum of files, as well as a timeline that allows zooming in on a particular event range to see if the number of files dropped unexpectedly.  The data include diverse experimental outputs from multiple instruments and detectors. For example, Figure~\ref{fig:stats_dashboard}(c) tracks as a function of time the number of scans acquired using two distinct diffraction techniques (far-field and near-field high-energy diffraction microscopy (HEDM), labeled "ff" and "nf" here), each performed on a separate detector.  During the deployment documented in this work, custom widgets were created to plot experiment-specific parameters, including mechanical load, position of critical motors, and specimen temperature.




Dashboards are customized for each user group, as illustrated in Figure \ref{fig:stats_dashboard}. Common elements include distributions of file types, aggregate data volumes, and timelines showing instantaneous and cumulative file production. A single beamtime can generate up to 100 terabytes of data across NeXus, HDF5, CBF, and TIFF formats. Visualizing these statistics helps users detect interruptions, slowdowns, or unexpected drops in file generation. Experiment-specific widgets can display parameters such as mechanical load, motor positions, and specimen temperature, supporting in situ studies.

The examples in Figure \ref{fig:beamline_timeline}, drawn from beamline ID3A, demonstrate how the dashboard reveals patterns during active experiments. File-type distributions help users anticipate storage needs, while beamline timelines correlate activity with data production. The far-field and near-field scan counts in Figure \ref{fig:far_and_near_field} illustrate how experiment structure evolves over time and assist users in verifying scan completeness. Together, these capabilities provide a real-time, data-driven view of beamline operations and significantly enhance experiment oversight.

\subsection{Data Evaluation and Visualization Dashboard}\label{sec:vis_dash}
 
The data acquisition dashboard is complemented by the data evaluation and visualization dashboard shown in Figure~\ref{fig:dash_explained}. Data evaluation dashboards are created dataset-by-dataset by placing JSON files in specific locations on the networked file system shared between the CHESS on-site network and the NSDF EntryPoint. Each JSON file contains a message for the dashboard application, which specifies the location of a dataset on the CHESS DAQ and the metadata to attach to that dataset. When the dashboard application receives a new message, it converts the specified data on the CHESS DAQ to an efficient format for streaming visualization on the NSDF EntryPoint and makes the new dataset available for visualization. These messages are generated immediately after data acquisition by tying into experiment control in SPEC, as described in Section~\ref{sec:entrypoint}. In this way, data evaluation dashboards are generated and populated through an automated process that is transparent to users.

Our dashboard is built upon open source repositories OpenVisus~\cite{Openvisus_git} and OpenVisusPy and includes basic visualization settings such as changing dataset, palette, resolution, log or linear colormap functional space, metadata, slice direction, slice depth, and range. OpenVisus reorganizes raw detector output into a pyramid of progressively downsampled levels, stored in small but spatially coherent blocks rather than single big  arrays. This design is essential for high-throughput synchrotron workloads as it avoids loading entire multi-gigabyte volumes into memory. Instead, the dashboard retrieves only the specific blocks needed for a slice, zoom level, or probe query. As a result, users can evaluate large datasets rapidly, even as acquisition is still underway. Regions of interest can be visualized before the full dataset has been written, and fine-grained tiles are fetched only when required, reducing I/O pressure and enabling fluid interaction with partial or incomplete datasets. This block-based design is what allows the system to support on-the-fly visualization for experiments that generate large arrays at high frame rates.

One of the first requests received from staff scientists at CHESS was a side panel that showed the data through the volume at a particular point, as if using a probe on a selected point or region in the volume.  This iteration led to our probe dashboard (Figures~\ref{fig:probe_with_missing_data} and~\ref{fig:dash_explained}). 
  Traditionally, such analysis required offline batch processing of full volumes. \textbf{By leveraging the  hierarchical layout of the data stored, the probe tool extracts only the blocks needed along a selected path or within a user-defined window, computing minimum, maximum, or mean values without reading the entire dataset.} This enables interactive comparison of multiple points of interest, allowing scientists to identify structural variations, missing slices, detector artifacts, or experimental anomalies in real time. The result is a workflow in which users can interactively explore fully three-dimensional datasets as they are being generated, significantly improving responsiveness and data-driven decision-making during beamtime.

\section{Deployments and Assessment}\label{sec:deploy_eval}
\vspace{-0.2mm}

Dashboards are deployed on the CHESS EntryPoint using Docker containers run by an NSDF service account. Developers and dashboard maintainers use their own private CHESS credentials (username and password) to access this service account and environment. From there, they can monitor, start, and stop existing dashboards and deploy new dashboards using sudo. In addition to deploying live dashboards, the service account creates static HTML pages displaying various experimental statistics served using an Apache HTTPD server.

\vspace{-0.2mm}

\subsection{Deploying at Beamlines}\label{sec:deploy_beamline}
The diverse techniques employed by different beamlines create unique requirements for their dashboards. Addressing these requirements is crucial for the dashboard to be useful for the particular beamline.

\noindent {\textbf{ The Forming and Shaping Technology Beamline (ID3A) at CHESS}  focuses on investigations into the relationships between processing, structure, properties, and performance in load-bearing materials (e.g., metals, cements, ceramics). In particular, ID3A specializes in in-situ imaging and diffraction measurements performed during thermomechanical loading and processing. Of the standard techniques at this beamline, micro-computed tomography and high energy diffraction micrscopy (HEDM) are particularly data-intensive, requiring specimen rotation through up to 360 degrees, with area detector images captured typically every tenth to quarter of a degree. These datasets are challenging to visualize in real-time at the beamline, historically relying on users opening large image stacks in software such as ImageJ or HEXRD. Opening a full rotation series dataset (hundreds to thousands of frames from a single sample rotation) on the beamline control computer risks disrupting ongoing data collection due to high memory demands. Consequently, users relied on separate compute resources to visually inspect their data. These checks are essential for experiment success, enabling fine-tuning of parameters and early detection of errors or instrumentation glitches that could otherwise compromise the dataset.

\textbf{Our data analytics dashboards represent a game-changing advancement for ID3A user groups, significantly lowering the barrier to data visualization and quality checks.} Our dashboards are highly effective in allowing multiple experimenters to examine raw data independently as collected without having to transfer files or interrupt work at the beamline control computer. As described in Section~\ref{sec:usecases_umich}, this allows for streamlined real-time collaboration within geographically distributed research teams.

\noindent{\textbf{ The Functional Materials Beamline (ID3B) at CHESS}} is dedicated to studying soft materials such as polymers, composites, and thin films using advanced X-ray techniques. It supports simultaneous SAXS/WAXS microscopy, GISAXS/GIWAXS, and phase-contrast imaging to probe properties at atomic and molecular scales. These methods enable real-time observation of internal structures and dynamic behaviors during processing or under variable conditions, offering critical insights for developing technologies in electronics, energy storage, and advanced manufacturing. By integrating state-of-the-art synchrotron methods with materials science challenges, ID3B accelerates the design and optimization of functional materials for specific applications.

A few user groups have successfully used the dashboards on this beamline, but large-scale adoption has not yet occurred due to higher-priority efforts. We continue working with collaborators to expand dashboard use at ID3B.

\noindent{\textbf{ The Quantum Materials Beamline (ID4B) at CHESS}} is a specialized facility designed to study complex materials with unique quantum properties that classical physics cannot explain. Using advanced X-ray techniques, researchers can probe the intricate interplay of electronic, magnetic, and structural properties in these materials at the atomic scale. The beamline is equipped with state-of-the-art instrumentation, including a high-resolution detector and cryogenic systems, allowing scientists to observe low-temperature quantum phenomena. By enabling detailed visualization and analysis of quantum materials, ID4B contributes to developing next-generation technologies in fields like quantum computing, advanced electronics, and energy storage. 

The quantum materials community, in particular, has experienced transformative benefits from the innovative dashboard at ID4B. {\textbf{ The deployed dashboards do more than enhance data visualization; they open up groundbreaking opportunities for in-depth analysis of quantum material datasets.}}

\subsection{Assessment of Our Framework}\label{sec:eval}
The dashboard framework’s effectiveness was evaluated during real-time data acquisition by CHESS beamline scientists and users. User feedback highlighted substantial improvements in accessibility, efficiency, and scientific productivity. We detail the assessment of remote monitoring, data acquisition, quality control, and data validation dashboards, followed by two beamline case studies.

Deployed across three beamlines (ID3A, ID3B, ID4B) at CHESS, the web-based dashboard has been extensively tested with beamline scientists and research scientists from 43 research groups. Over the last three months of 2024 alone, the dashboards managed 50–100 terabytes of data and over 10 million files, demonstrating the scalability and effectiveness of our framework. {\textbf{ Researchers used the dashboards to monitor their experiments remotely, validate data quality in real-time, and optimize their workflows. } }

\subsubsection{Assessment: Remote Monitoring}\label{sec:eval_remote}
The dashboard framework significantly improved remote access and monitoring at CHESS. Previously, remote users depended on the facility status website for major alerts, the NoMachine Enterprise Client to view on-site desktops, and constant communication with collaborators to track experiment progress. NoMachine requires interacting with a full remote Linux desktop, with repeated logins and sensitivity to network latency. Users often juggle multiple terminals, scripts, and GUIs across several instruments, making this workflow cumbersome for remote participants who must locate and manage shifting windows. Without active communication with on-site collaborators, reviewing data in real time was often difficult.

The new framework enables direct access to live data streams from any device without logging into beamline machines, significantly reducing logistical overhead. As one staff scientist summarized, “You can access data from anywhere on any device like never before,” capturing the improved flexibility and usability provided by our system.

\subsubsection{Assessment: Data Quality Control}\label{sec:eval_quality}
Handling terabytes of data makes real-time quality assurance a persistent challenge. Although automated scripts manage data collection, they still require oversight to catch issues early. The web-based acquisition statistics dashboard enables “invaluable monitoring” without repeated logins to NoMachine or CHESS clusters. Staff scientists can now track acquisition status, data reduction, and group progress directly, improving operational efficiency and helping maintain data quality throughout collection. This capability establishes a new standard for beamline data handling by providing seamless access from any device. As one scientist noted, “Up to now, CHESS lacked standardized visualization tools: our tools were hard to access remotely, cumbersome to operate, with poor scaling for large datasets. Sometimes, data collection errors were not discovered until after CHESS beamtime. NSDF dashboards allow experimenters to use beamtime more effectively.”

\subsubsection{Assessment: Data Acquisition }\label{sec:eval_acq}
The statistics dashboards had an unexpected benefit for large data acquisition. Beamline users face the practical challenge of monitoring ongoing scans 24 hours per day during beamtime. The statistics dashboards give beamline users a view into the data collection process. They identify times when collections have started or completed and when the beam has gone down due to facilty-wide issues. This helps identify errors during collection, helps schedule in-person time, and allows for estimation of task completion. Additionally, real-time monitoring gives researchers a view into data collection sizes.

\begin{figure}[t!]
    \centering
    \includegraphics[width=0.95\linewidth]{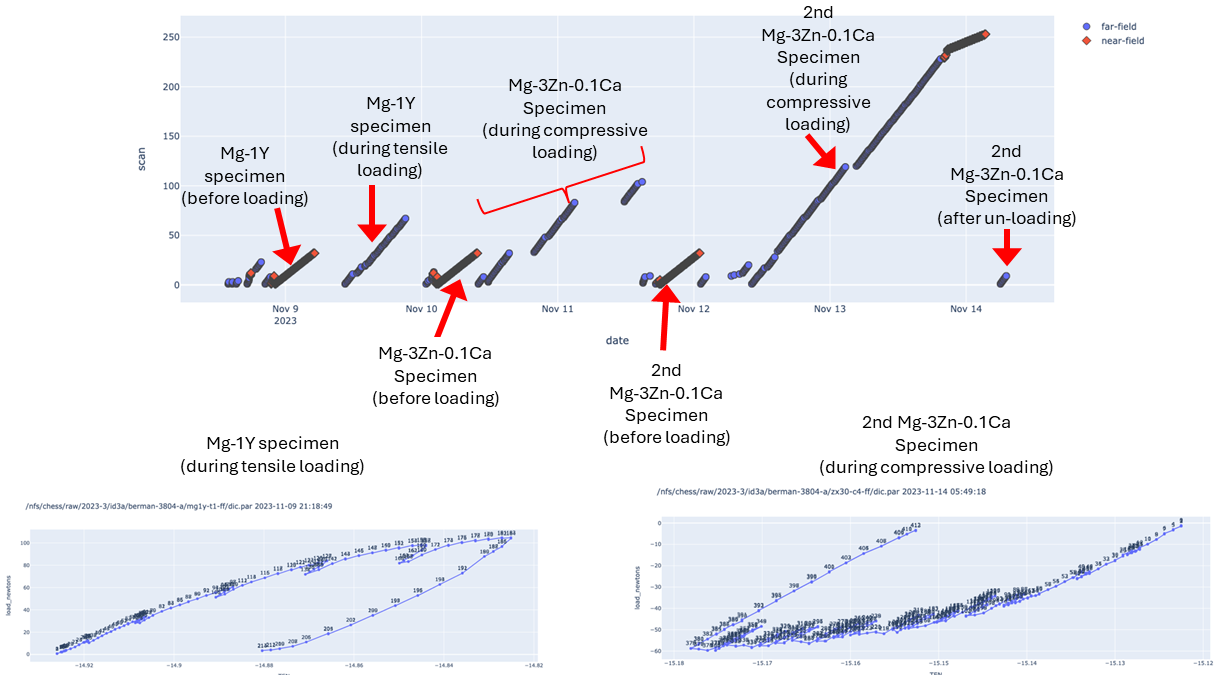}
    \caption{Top: plot of scan number as a function of time. Annotations have been added to aid in interpretation. Bottom: Plots using our stats dashboard of mechanical load versus tension motor position for tensile loading of specimen Mg-1Y (left) and compressive loading of specimen Mg-3Zn-0.1Ca (right).  
  }
    \label{fig:stressStrainPlot}
        \vspace{-5mm}
\end{figure}

\subsubsection{Assessment: Data Evaluation and Visualization}\label{sec:eval_vis} 
The large data validation dashboard enables immediate access to acquired data, significantly reducing the time required for data review and decision-making. With real-time visualization and interactive image profiling capabilities, researchers can quickly assess localized behaviors and inspect image features at specific points of interest over time. This immediate feedback loop enables on-the-fly corrections during experimental runs, promptly addressing any issues, as detailed in Section~\ref{sec:usecases}.



\subsection{Deployment Use Cases}\label{sec:usecases}
In this section we document two use cases that provide domain expert assessment and validation on two different beamlines.  The first offers feedback from ID3A, focusing on a representative user experiment centered on examining the mechanical behavior of magnesium alloys.  The second use case provides feedback from CHESS staff scientists at ID4B.

\subsubsection{Use Case 1: Forming and Shaping Technology Beamline (ID3A)}\label{sec:usecases_umich} 
At ID3A, the dashboards have supported 11 user groups conducting micro-computed tomography, powder diffraction, and near- and far-field HEDM (nf-HEDM and ff-HEDM). The first deployment in November 2023 coincided with a University of Michigan beamtime, led by developers of the Materials Commons repository~\cite{MaterialsCommons2025}, focused on deformation mechanisms in two magnesium alloys with potential for lightweight automotive applications. nf-HEDM scans characterized grain morphology and crystallographic orientation in undeformed specimens, while ff-HEDM scans captured grain-average elastic strain, orientation, and position before and during loading. Comparing undeformed and deformed states revealed changes in diffraction patterns, offering insights into active deformation mechanisms.

Because the principal scientist could not travel due to a recent and unplanned surgery, collaborators relied on the dashboards for remote monitoring. Unlike the CHESS facility status site, which only reports beamline health, our dashboards provide direct insight into data quality. Accessible from any internet-enabled device, they allow users to verify acquisition progress and review data statistics in real time.
The data acquisition dashboard (Section~\ref{sec:dataAcq}) enables efficient offsite monitoring by giving users a quick visual overview of experiment progress, even during brief breaks.

\begin{figure}[t!]
    \centering
    \includegraphics[width=0.9\linewidth]{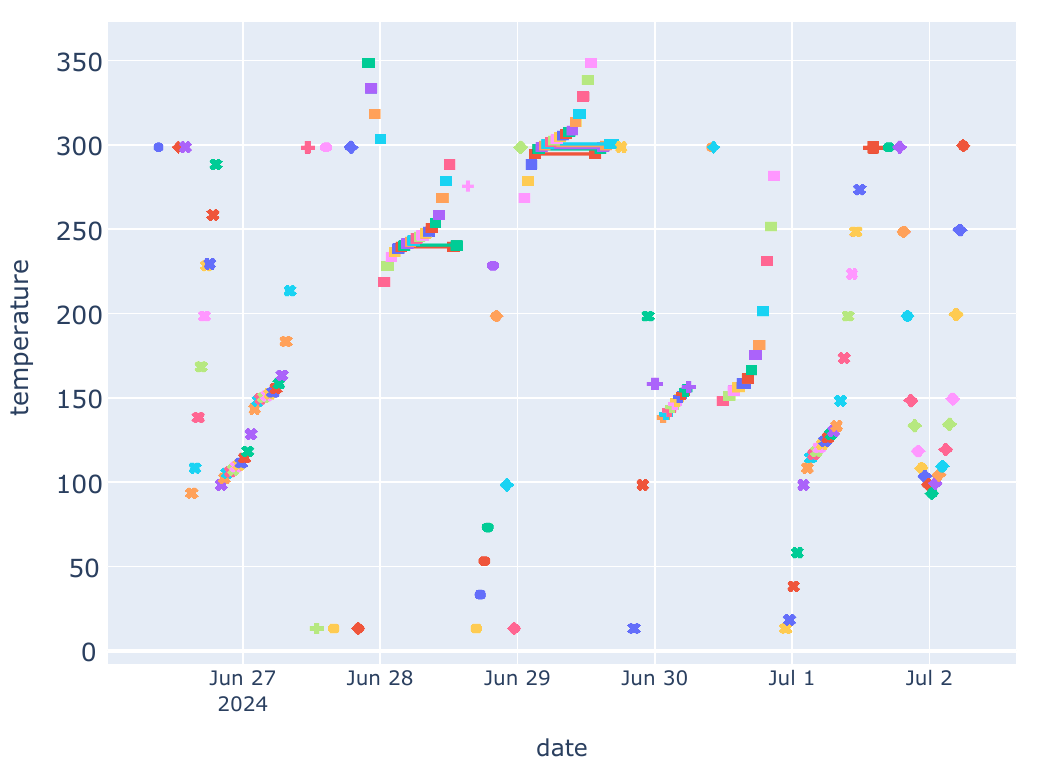}
    \caption{Dashboard widget for the beamline ID4B Quantum Materials (QM2) provide real-times plot for temperature dependent scan of different samples (shown above) as well as visualization of the data (in supplemental materials).  
}

    \label{fig:QM2}
     \vspace{-5mm}
\end{figure}

The experiment timeline, shown in Figures~\ref{fig:timing_down} and~\ref{fig:beamline_timeline}, outlines the overall beamtime, illustrating how time was spent and when users switched between techniques. It also provides a quick reference for interruptions, such as the beam going down at 3:30 am on November 14th, 2023. While the CHESS facility status website reports beam conditions, the acquisition dashboard statistics let remote users see exactly during which scan the beam was lost.

As mentioned in Section~\ref{sec:collab}, another unique and critical feature of our collaborative deployment was the adaptability to customize dashboards for individual experiments. Once a mechanical test was initiated, plots of mechanical load versus load frame motor position were automatically created and updated in the dashboard (Figure~\ref{fig:stressStrainPlot}). 
Examining the raw detector images in the unloaded nf-HEDM scans via the data evaluation dashboard was helpful in assessing the quality of the data, shown in Figure~\ref{fig:dash_explained}. The size of individual grains within a specimen is a critical contributing factor to the quality of both nf- and ff-HEDM data; grains that fall below a certain threshold relative to the incident X-ray beam size are often difficult or impossible to reconstruct from the raw diffraction data. Although one of the materials in this experiment had smaller grains than desired, \textbf{the visualization tools within the data evaluation dashboard helped the researchers determine, in real time, that the data was of sufficient quality to yield useful results}. 

User feedback emphasized the value of integrating high-rate acquisition data with experimental metadata. As the University of Michigan team noted, “stress-strain plots... are a nice quick reference for how a loading experiment is progressing… [and] the stats from the [data acquisition] dashboard are a nice complement to this information. A link to the Stats page and visualization dashboard would help enrich a published HEDM dataset.” These feedback present the importance of unifying streaming data, metadata, and analysis views for data-intensive experimental workflows.

\subsubsection{Use Case 2: Quantum Materials Beamline (ID4B)}\label{sec:usecases_chess_scientist}  
Since the launch of the dashboard’s status page, the ID4B QM2 beamline has tracked its first-ever real-time data collection. It also allowed the monitoring of data analysis status, even when users performed beamline-specific data analysis from their home institutions.   
This has proven tremendously beneficial for the beamline scientists, allowing them to keep track of the final data analysis status for about 22 different user groups at QM2, both nationally and internationally. 
This enhancement is essential for driving impactful scientific discoveries. The dashboards include a web-based HDF5 file viewer, which significantly enhances data exploration, shown in Figure~\ref{fig:QM2}. This feature is particularly valuable for the QM2 beamline, where traditional access methods have been challenging, and the web-based approach offers streamlined access to HDF5 files. Traditionally, accessing QM2 data through NeXpy requires navigating a specific environment, often posing user challenges. {\textbf{ In contrast, the web-based interface streamlines this process, allowing users to explore complex data files easily.}} 



 \subsection{On-going developments}
We continue to refine our understanding of staff and user needs at each beamline and iteratively improve the dashboards. Our initial focus on rapid deployment limited metadata collection on user engagement, such as number of users, experiment duration, or access patterns. We are now adding this tracking, along with mechanisms for users to capture “insight moments” during exploration to support journaling and reproducibility. Future work includes widgets for comparing data volumes across beamtimes, AI tools for anomaly detection and downtime analysis, and integration with the CHESS data management framework for long-term preservation.

We hope this paper encourages broader adoption of our tools by beamline users and staff scientists, including at CHESS’s Functional Materials Beamline (ID3B) and other synchrotron facilities. \textbf{Drawing on lessons from this deployment, we are collaborating with scientists at Oak Ridge National Laboratory (ORNL) to extend these interfaces to their beamlines.} We are also developing new dashboards, such as the one now being deployed (Figure~\ref{fig:disc_probe}). An effective viewer for ff-HEDM data remains a work in progress and may require innovative approaches to distill large, sparsely populated pixel arrays into a more intuitive format.

\begin{figure}[t!]
    \centering
    \includegraphics[width=0.9\linewidth]{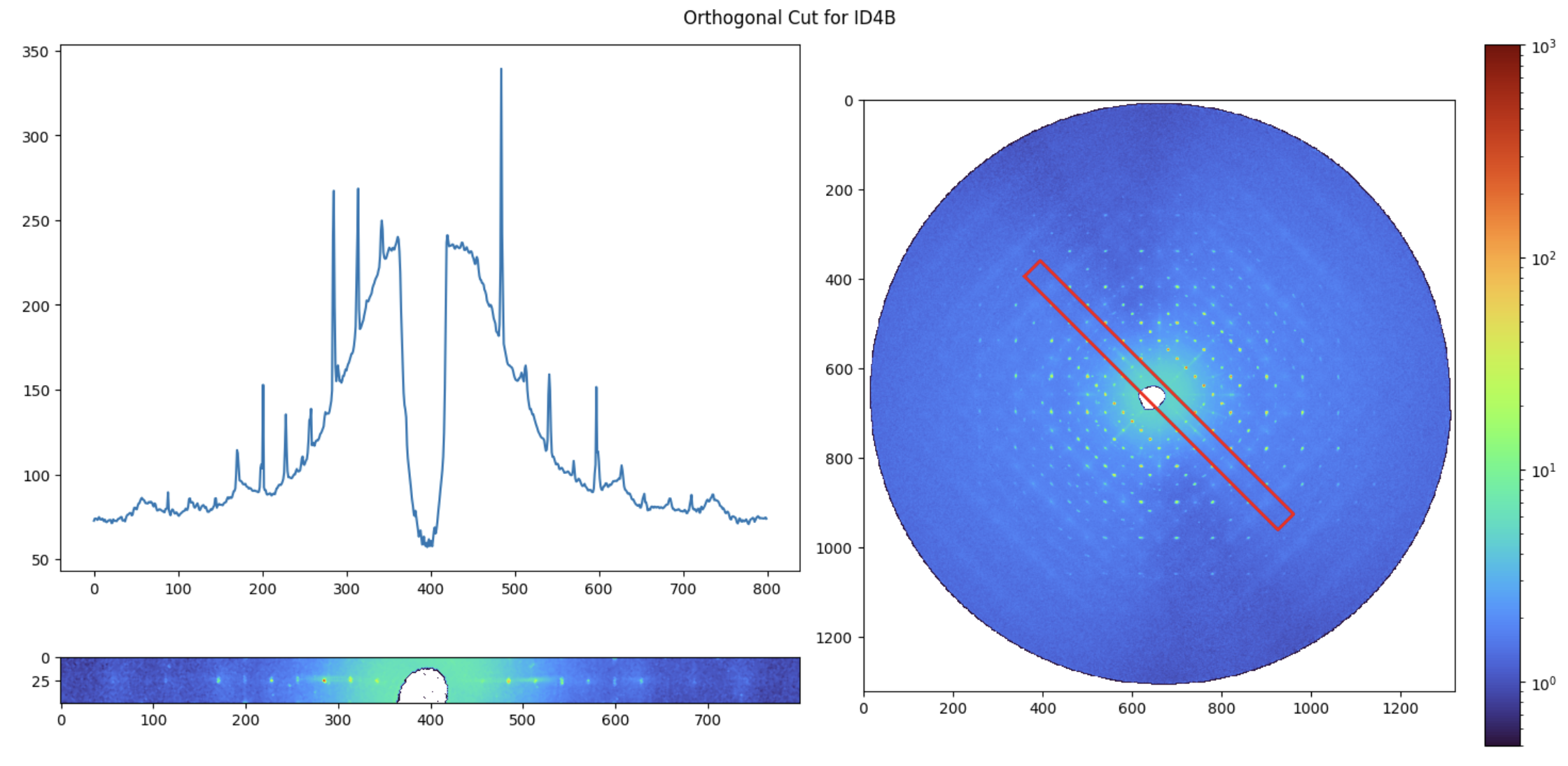}
    \caption{ Extracting an orthogonal slice from a 2D image of ID4B beamline data from our Jupyter notebooks. The top-left plot shows the summed values across each column in the cropped region, the bottom-left image displays the cropped region using a log-scale color map, and the right panel shows the original image with the bounding rectangle (red) used for cropping.
}
    \label{fig:disc_probe}
        \vspace{-6mm}
\end{figure}

\vspace{-0.5mm}
\section{Outcomes and Conclusion}
\vspace{-0.5mm}

The deployment of the dashboard framework at CHESS resulted in several significant outcomes for large-scale data acquisition and analysis.  
Real-time acquisition and visualization of terabyte-scale datasets are often hampered by downtime, data loss, and lengthy validation delays. Our framework addresses these challenges by integrating acquisition statistics and visualization into web-based dashboards, giving staff and users up-to-date views of ongoing experiments without downloading data or being physically present. This has optimized use of limited beamtime and improved overall workflow efficiency.  Customized dashboards for individual beamlines demonstrate the adaptability of our system to diverse techniques, data rates, and pipelines at different facilities. Ongoing bi-weekly iterations expand features and automation, with positive feedback confirming its impact and potential for broader adoption. By focusing on essential parameters for each user group, we streamline acquisition processes while laying the foundation for advanced analytics.

While large-scale facilities generate massive datasets, few solutions make them accessible and actionable in real time. Our framework offers a scalable, end-to-end approach that connects data acquisition to scientific insight, providing a model for data-driven discovery at scale. Moreover, the challenges addressed at CHESS are common across synchrotron and neutron facilities, positioning this framework as a transferable technology that can be adopted broadly to enhance scientific productivity worldwide.

\section{Acknowledgements}
 The work is supported in part by NSF OAC award 2138811, NSF 
CI CoE Award 2127548, DOE SBIR Phase I DE-SC0026005, DOE SBIR Phase II DE-SC0017152, NSF POSE Phase I Award 2449026, NSF OISE award 2330582, the Advanced Research Projects Agency for Health (ARPA-H) grant no. D24AC00338-00, the Intel oneAPI Centers of Excellence at University of Utah, the NASA AMES cooperative agreements 80NSSC23M0013 and NASA JPL Subcontract No. 1685389. This paper is also supported by the WIRED Global Center, jointly funded by the NSF GC grant 2330582 and NSERC grant ALLRP 585094-23.
This work is also based on research conducted at the Center for High-Energy X-ray Sciences (CHEXS), which is supported by the National Science Foundation (BIO, ENG and MPS Directorates) under award DMR-2342336. This material is based on research sponsored by AFRL under agreement number FA8650-22-2-5200. The U.S. Government is authorized to reproduce and distribute reprints for Governmental purposes notwithstanding any copyright notation thereon. This work was supported, in part, by the U.S. Department of Energy, Office of Basic Energy Sciences, Division of Materials Sciences and Engineering under Award DE-SC0008637 as part of the Center for PRedictive Integrated Structural Materials Science (PRISMS Center) at University of Michigan.  

\bibliographystyle{IEEEtran}
\bibliography{template.bib}

\end{document}